\documentclass[preprint,aps]{revtex4}

\usepackage{graphicx}
\usepackage{dcolumn}
\usepackage{bm}

\begin{document}


\title{On tight multiparty Bell inequalities for many settings}

\author{Marek \.{Z}ukowski}

\affiliation{
 Institut f\"{u}r Experimentalphysik, Universit\"{a}t Wien,
Boltzmanngasse 5, A-1090, Wien, Austria
\\
Instytut Fizyki
Teoretycznej i Astrofizyki Uniwersytet Gda\'{n}ski, PL-80-952
Gda\'{n}sk, Poland\\
Tsinghua University, Beijing, China}

\date{\today}

\begin{abstract}{A derivation method 
is given which leads to a series of tight Bell inequalities for experiments involving $N$ parties, with  binary observables, and three
possible local settings. The approach can be generalized to more settings. Ramifications are presented. }
\end{abstract}

\maketitle

\section{Introduction}
This paper is not devoted to the meaning and ramifications of Bell's theorem. Rather, it will study some technical aspects associated with multiparty Bell
inequalities. Multiparty Bell inequalities are recently gaining in importance, because security, or performance, of many of quantum communication schemes
(like multiparty key  (secret) sharing \cite{SCARANIGISIN}, quantum communication complexity problems \cite{BZPZ}) can be measured with Bell inequalities of some form \cite{USEFUL_ENT}. Bell inequalities show the limit of achievable correlations between many parties, if they share only classical means of communication and/or 
computation. If violated by certain predictions for a certain  entangled state, they show that this state, without any further manipulations is a good resource for some quantum informational tasks. To put is short, in the standard context Bell inequalities show that if there are $N$ separated parties, each equipped with a classical supercomputer, who share common data, and perhaps programs, but otherwise are not allowed to communicate, the locally run programs cannot 
simulate some phenomena associated with spatially separated $N$ local measurements, with locally decided settings, on $N$ entangled particles, when each party is asked to predict (compute) result of a measurement for a different particle.

Bell's theorem was formulated for two particles \cite{BELL64}. It took a quarter of century to realize that for more than two particles the situation is much more interesting. The discovery of Greenberger-Horne-Zeilinger correlations \cite{GHZ} immediately led to various generalizations of the multipartite problem. 
This gave birth to the quest of finding Bell inequalities for the new problems. Mermin was first to produce series of inequalities for arbitrary many particles, involving dichotomic observables, and allowing each observer to choose between two settings \cite{MERMIN}.
A complementary series of inequalities was introduced by Ardehali \cite{ARDEHALI}. In the next step Belinskii and Klyshko gave series of two settings inequalities, which contained the tight inequalities of Mermin and Ardehali \cite{BELINSKII-KLYSHKO}. Finally the full set of tight two setting per observer, $N$ party Bell inequalities for dichotomic observables was found independently in \cite{WW, WZ,ZB}. All these series of inequalities
are a generalization of the CHSH ones \cite{CHSH}. Such inequalities involve only $N$ party correlation functions. The aim of this paper is to go one step further, and to produce series of tight CHSH-type Bell inequalities for $N$ observers, dichotomic observables, and {\em{three}} possible setting for each observer. Recently series of tight inequalities, which do not form a complete set, were found for the case when $k$-th partner can  choose between $2^{k-1}$ settings (see \cite{WUZONG1, WUZONG2, LPZB}). We shall not follow this approach here, as it seems to be incapable of generating
the full set of Bell inequalities. Also the methods of \cite{KWEK} and \cite{ROT} will be not discussed here, as they do not follow from the analysis of the geometry of the polytope of local realistic models (for this concept see \cite{PITOVSKY}). 

Only the proof for the necessary condition for the two-observer ($3\times3$) case will be shown in full detail. As the reader will see, the generalization 
to more parties, that is to $3\times3\cdots\times3$ problems, is straightforward. In all these cases we shall be concerned with the generalization of the inequalities of the CHSH type. Such inequalities involve only $N$ party correlation functions.
However, a remark will be made on how to extend the validity of the presented results to CH-type inequalities \cite{CH} (which involve also correlations of lower rank). We shall study only experiments involving dichotomic observables.

We have chosen the three settings case, with two valued observables, because
it is the simplest example for which the method used here can lead to new results.

\section{The $3\times3$ problem}

We shall try to derive Bell inequalities for the case
two-particle correlation function. 
We allow each 
observer to choose between three settings. We will search for the positivity 
condition for the local realistic (unphysical, hidden) probability distribution involving results for all settings:
\begin{equation}
P(m_1^{(0)},m_1^{(1)},m_1^{(2)},m_2^{(0)},m_2^{(1)},m_2^{(2)}),
\end{equation}  
where $m_i^{(n_i)}$ is the local realistic prediction for measurement result
if the observer $i$ chooses to measure a dichotomic observable number $n_i$. Note, that due to complementarity or, formally, non-commutability, of the algebra of observables such distributions are impossible within the quantum mechanical formalism.
  
Let us introduce the following new dichotomic variables:
\begin{eqnarray}
m_1^{(0)}m_2^{(0)}=x,\\
m_1^{(0)}m_1^{(1)}=a,\\
m_1^{(0)}m_1^{(2)}=b,\\
m_2^{(0)}m_2^{(1)}=c,\\
m_2^{(0)}m_2^{(2)}=d.\\ \nonumber
\end{eqnarray}
All these variables are independent of each other. 
That is, fixing four of them does 
not determine the fifth. 
However, as we shall see only these variables enter to the problem of finding the Bell inequalities involving correlation functions. 
That is our task is to find conditions for existence of
the positive distribution $P(x,a,b,c,d)$.

One can introduce an important technical tool. this will be the notion of a ``sign" function. 
We shall use throughout  
``sign" functions, $s$,  of four the dichotomic parameters $a,b,c, d=\pm1$. The sign function is dichotomic itself, that is we require $s(a,b,c,d)=\pm1$. 
The most general sign function of $a,b,c,d=\pm1$ has the following discrete multivariate Fourier decomposition:
\begin{eqnarray}
&s(a,b,c,d)=g_{00}+g_{01}c+g_{02}d+g_{10}a+g_{11}ac+g_{12}ad+g_{20}b+g_{21}bc+g_{22}bd& \nonumber \\
&+r_{1;0}ab+r_{1;1}abc+r_{1;2}abd + r_{2;0}cd+ r_{2;1}cda+ r_{2;2}cdb
+r_Dabcd\big).&\nonumber \\
\label{SIGNFUNCTIONS}
\end{eqnarray}

It will be shown in the next section that a specific family of sign functions, 
which have the following Fourier
decomposition
\begin{eqnarray}
&S(a,b,c,d)& \nonumber \\
&=g_{00}+g_{01}c+g_{02}d+g_{10}a+g_{11}ac+g_{12}ad+g_{20}b+g_{21}bc+g_{22}bd,& 
\nonumber \\
 \label{GOOD-SIGNFUNCTIONS}
\end{eqnarray}
plays an important role in the definition of the Bell inequalities for the problem.
Note that such functions simply do not involve products $ab$ and $cd$.
Each such a function leads to a Bell inequality of the form
\begin{eqnarray}
&|\sum_{n_1,n_2}g_{n_1n_2}E_{n_1n_2}|\leq2^4.&\label{B}
\end{eqnarray}
{\em This set of Bell inequalities, 
forming the necessary condition for the positivity of the local realistic 
distribution, can be
written as}
\begin{eqnarray}
&2^4+
\sum_{a,b,c,d=\pm1}S(a,b,c,d)\big(E_{00}+E_{01}c+E_{02}d+E_{10}a& \nonumber \\
&+E_{11}ac+
E_{12}ad+E_{20}b+E_{21}bc+E_{22}bd\big)\geq 0.&\nonumber \\
\label{POSITIVITY-SUM-1}
\end{eqnarray}

To find the explicit coefficients of the
 Bell inequalities, as it was said earlier, 
one must know the full family of the ``admissible" sign functions satisfying
(\ref{GOOD-SIGNFUNCTIONS}). For the $3\times3$ case it is elementary to 
show that there is a subfamily of 18 such functions, 
which are factorable, and they lead to trivial inequalities like $-1\leq E_{n_1n_2}\leq 1$.
There also is another sub family of non-factorable ones like 
$S(a,b,c,d)=S(a,b)=\pm\frac{1}{2}\big(1+a+c-ac\big)$, and all other ones that can be obtained from these ones by the permutation of $a$ and $b$, and by putting the minus sign in a different position, 
which lead to the good old CHSH inequalities. No other two-particle correlation function inequalities 
can de derived with this method. However, it was shown by Garg \cite{GARG} that the two observers three-settings-per-observer problem set of Bell inequalities for correlation functions contains only standard, $2\times2$,  CHSH inequalities. One has to remark here that, most importantly, the generalization of theses inequalities
to more than two parties leads to new inequalities, which involve three settings for some of the observers \cite{WIESNIAK-ZUK}. That is  we enter a new unexplored territory. 

\section{Derivation of necessary condition}
In this section we shall seek a necessary  condition for  
the existence of a local realistic model of the set of two-particle
 correlation functions for the 
Bell problem described above. The sufficiency proof of this condition is a bit lengthy and will be given elsewhere. 
We assume that the full set of correlation functions for the problem, $E_{n_1n_2}$ ,
is described by a single local realistic model. Obviously if such a model 
exists, then it is generated by
the hidden unphysical probability distribution 
discussed in the previous section.

\subsection{Geometrization of the problem}
One can write  down the full set of the values of the two 
particle correlation functions in the 
form of a nine dimensional vector, namely
$$(E_{00}, E_{01},E_{02}, E_{10},...,E_{22}).$$
It  belongs to a  nine dimensional real vector space $\bf{R}^9$.
Since the space is three times three dimensional, 
one can treat it as a set of tensors (three times three matrices), that is 
$\bf{R}^3\otimes \bf{R}^3$. That is,
 the correlation functions for all the $3\times3$ settings form
 a tensor $\hat{E}$ with 
components $E_{n_1n_2}$.

\subsubsection{Overcomplete bases}
The following set of tensor products, 
namely certain subsets of sixteen vectors of the set of vectors given by
$\pm(1,a,b)\otimes(1,c,d)$, with $a,b,c,d=\pm1$ form an overcomplete basis in a nine dimensional 
real vector space.
We can fix such an overcomplete basis, by fixing the sign in front of 
every such vector. The vectors belonging to any such basis 
can be written down in the following way
\begin{equation}
s(a,b,c,d)(1,a,b)\otimes(1,c,d)=V_{abcd,s}.
\label{VERTICES}\end{equation}
Note that each overcomplete basis of this family is defined by  a specific choice of the sign function $s(a,b,c,d)$.
It is easy to show, that if one treats the vectors  
as column matrices
then for each choice of the overcomplete set, or equivalently the function $s$,
 one has
\begin{equation}
\hat{I}=\frac{1}{2^4}\sum_{a,b,c,d=\pm1}V_{abcd,s}V_{abcd,s}^T,
\label{UNIT}
\end{equation}
where $\hat{I}$ is the unit operator in the space. The best way to check this formula is to
derive first its equivalent for the $\bf{R}^3$ space, for the ovecomplete set given by $(1,a,b)$.

\subsubsection{Local realistic correlation functions}

{\it If the full set of the nine 
correlation functions, $\hat{E}$ 
is to have a common local realistic model, 
it must be representable by a linear combination of the form
\begin{equation}\hat{E}=\sum_{a,b,c,d=\pm1}C_{abcd}V_{abcd,S},
\label{LOC-REAL}
\end{equation}
where 
the coefficients fulfill the following relation:
$$\sum_{{a,b,c,d}=\pm1}|C_{abcd}|\leq1.$$}

Such a statement is correct because our situation has some specific 
traits. Note, that the obvious condition to be satisfied is 
\begin{equation}
\hat{E}=\sum_{a,b,c,d=\pm1}\sum_{x=\pm1}P_{x,a,b,c,d}V_{abcd,x}.
\label{STANDARD}
\end{equation}
However it can be always reduced to the condition (\ref{LOC-REAL}), 
because the vectors 
$ V_{abcd,x}$, with $x=1$ and $x=-1$, point in exactly opposite directions.
{\it The local realistic interpretation of all that is that the vectors $V_{abcd,x}$ represent the full set of possible 
``deterministic orders" sent by the source to the observers (with probability 
$P({x,a,b,c,d})$)}.

\subsection{Necessary condition}
Let us derive a necessary condition for the existence of a local realistic 
model of the full set of the correlation functions. 
The derivation will not be based on probability 
theory, but rather on geometry of convex  polytopes. This is 
because the full set of $E$'s,
which have a local realistic model, that is satisfying (\ref{STANDARD}), 
is a convex polytope \cite{PITOVSKY}, with generating set, or set of $32$ ``vertices",
being the full set of the product tensors representing the possible 
deterministic orders ``sent" by the source 
to the two observers, that is (\ref{VERTICES}). 

It is well known that overcomplete bases have the following property.
The expansion coefficients in terms of such a basis uniquely define a 
vector in space. However, there exists infinitely many equivalent expansions
(of every vector)
in terms of the same overcomplete set.

Let us therefore select one specific expansion, which will play 
an important role in the sequel.
We shall call it ``canonical" with respect to a given overcomplete set.
In the present case of a specific  
overcomplete set, given by (\ref{VERTICES}),  
 it is defined in the following way:
If a vector, $V$, belongs to our space (here, nine dimensional) 
it can be represented as
\begin{equation}
V= \sum_{a,b,c,d=\pm1}q(a,b,c,d,s)V_{abcd,s},
\label{CANONICAL}
\end{equation}
where 
the expansion coefficients are given by the ``natural" formula
\begin{equation}
q(a,b,c,d,s)=\frac{1}{2^4}V_{abcd,s}\cdot V.
\end{equation}
By $A\cdot B$ is denoted the scalar product in the nine dimensional space.
To see that this definition is consistent, 
it is enough to look back at the expansion of the unit operator, 
(\ref{UNIT}). 

{\em Lemma}. Assume  that 
the tensor $\hat{E}$ is representable in the form of a convex combination of all 
factorable tensors $$ V_{abcd,x}=x(1,a,b)\otimes(1,c,d),$$
where $x=\pm1$. 
Therefore the following condition holds:
If one takes the unit operator in the form
$$I=\frac{1}{4}\sum_{a,b,c,d=\pm1}V_{abcd,S}V_{abcd,S}^T,$$
then the sum of canonical 
expansion coefficients of $E$ in terms of the specific 
overcomplete set, 
defined by the sign function $S(a,b,c,d)$,  for any sign function $S$ of the admissible form
(\ref{GOOD-SIGNFUNCTIONS}),
satisfies the following conditions
\begin{equation}
-1\leq\sum_{a,b,c,d=\pm1}q(a,b,c,d,S)\leq 1.
\label{BELL-11}
\end{equation}

Note that
\begin{eqnarray}
&q(a,b,c,d,S)_E=\hat{E}\cdot V_{abcd,S} =\frac{1}{2^4}S(a,b,c,d)\big(E_{00}+E_{01}c+E_{02}d+E_{10}a& \nonumber \\
&+E_{11}ac+
E_{12}ad+E_{20}b+E_{21}bc+E_{22}bd\big)&\nonumber \\
\end{eqnarray}

{\em Proof}. One can show that every vertex of the convex  polytope, 
$V_{abcd,x}$, saturates the 
left or right hand side inequality 
in (\ref{BELL-11})involving the sign admissible sign function $S$. Since the inequalities are linear, therefore they must 
be satisfied by any convex combination of the vertices. 

More importantly, every right hand side inequality in (\ref{BELL-11}) is 
saturated by half of all vertices, namely all vertices $V_{abcd,S}$, whereas the left hand side inequality is satisfied by all vertices 
$V_{abcd,-S}$. Since each of these sets of vertices contains 16 elements, but is must contain 9 elements which are linearly independent, because otherwise it would not have been (over-)complete.  Because of that each inequality defines a face of the polytope (a face must contain  at least 9 vertices, which satisfy an equality, $E(V)=0$, defining a hyperplane which they span; all  vertices, $V'$, which do not belong to the face, when  inserted into the equality must then give $E(V')$ which is non-zero, and always of the same sign).

Therefore the presented Bell inequalities hold, and are tight (each defines a face of the convex polytope of local realistic models of the problem).     

\section{Hint about sufficiency proof \cite{ZUK}}
The proof, that the described set of Bell inequalities is the complete set of 
tight inequalities for the problem, will be presented elsewhere. It boils down to showing that
only the sets of vertices $V{abcd,S}$ defined my the admissible sign functions $S$, given by (\ref{GOOD-SIGNFUNCTIONS}),
can be in a single face of the polytope. In other words, each admissible sign function defines a face. No other faces exist.
Thus the set of inequalities must be complete.

\section{Ramifications}
The procedure 
can be generalized to an arbitrary Bell problem involving
$N$ parties, two valued observables, and $3$ local settings per observer. Generalization 
to more than two parties is trivial, and will be presented below.
So is generalization to more than three settings of the derivation of necessary conditions (sufficiency proof is not easy anymore). 

\subsection{Three or more observers}
For $N>2$ it is straightforward 
to write down the full set of tight Bell inequalities in the form of a single ``synthetic" formula.
E.g., for three parties, dichotomic observables, and three 
settings at each side such a set of tight Bell inequalities is given by
\begin{equation}
-1\leq\sum_{a,b,c,d,e,f=\pm1}q^{[3]}(a,b,c,d,e,f,S^{[3]})\leq1,
\end{equation} 
where 
\begin{itemize}
\item
$S^{[3]}$ is any admissible ``sign" function of the variables $a,b,c,d,e,f$ 
with the property that in its Fourier expansion products $ab$, $cd$ and $ef$
never appear (i.e., all products of indices of a local observer are missing),
\item the following overcomplete bases are selected from the set of 
deterministic 
``orders" sent by the local realistic source to the observers:
\begin{equation}
S^{[3]}(a,b,c,d,e,f)(1,a,b)\otimes(1,c,d)\otimes(1,e,f)=
V^{[3]}_{abcdef,S^{[3]}},
\label{VERTICES-3}\end{equation}
where most importantly the sign function belongs to the class of admissible ones,
\item
and finally 
\begin{equation}
q^{[3]}(a,b,c,d,e,f,S^{[3]})=\hat{E}\cdot V^{[3]}_{abcdef,S^{[3]}},
\label{BELL-GHZ}
\end{equation} 
where the set of three party correlation functions 
$E_{000}, E_{001},...E_{222}$ is written down as a vector (or rather a  three index tensor, $\hat{E}$) in 
$\bf{R}^{27}=\bf{R}^3\otimes\bf{R}^3\otimes\bf{R}^3$.
\end{itemize}
Note that the set of inequalities (\ref{BELL-GHZ}) has never 
appeared in the literature, just like zillions of other ones derivable by this method.

\subsection{Other generalizations}

If one wants to build Bell inequalities involving the full set of 
observable data for the experiment, that is
in the two parties case, also local averages of the local results, 
not only the 
correlation functions of products
results of the two parties, one can re-interpret the inequalities in the following way.

Let us consider the two-observer, {\em two-settings} problem, with dichotomic observables.
Consider the following subset of the vertices of the $3\times3$
problem
\begin{equation}
W'(m_1^{(1)}m_1^{(2)};m_2^{(1)}m_2^{(2)})=(1,m_1^{(1)},m_1^{(2)})
\otimes(1,m_2^{(1)},m_2^{(2)}).
\label{NEWTENSOR}
\end{equation}
Simply the hiddden results of the local experiments for the first settings, $m_j^{0}$ with $j=1,2$, was repalaced by $1$. 
Now the averages of components of the (\ref{NEWTENSOR}) of such vercites (or if you like convex combinations of all of them), that is
\begin{equation}
\sum_{ W'}P(W')W',
\label{MODEL-W}
\end{equation}
give: normalization condition,  averages of the local results, and finally all correlation functions, for all combinations of the two pairs of settings. I.e. nothing is missing 
from the full description of the phenomena, for the problem in question. Thus we get the full description of observable phenomena for an experiment
for which a local realistic model exist. Since the new vertices satisfy the inequalities derived for the $3\times3$ problem 
involving correlation functions, we get immediately a necessary condition for the existence of (\ref{MODEL-W}).  In other words, we get the CH inequalities. When this approach is (straightforwardly) generalized to more than two parties new families of tight Bell inequalities can be derived.

\subsection{Relation with the complete set of two settings inequalities}

The complete set of two setting Bell inequalities for $N$
parties, dichotomic observables and two possible settings for each party, can be recovered from the family of inequalities presented here.
Consider first the two party problem. One can  choose the subset of admissible sign functions of the following form $S(a,b,c,d)=S(a,c)$, i.e. depending on only one index per observer. Due to this fact all components of $S$ in the expansion (\ref{GOOD-SIGNFUNCTIONS}) which contain $b$ and $d$ drop out, and the inequality reduces to a two-setting one.
Similarly, for three parties problem one can choose
$S^{[3]}(a,b,c,d,e,f)=S^{[3]}(a,c,e),$
 with equivalent result, etc. 
 
\section{Acknowledgements}
Author thanks Anton Zeiliger for our first 15 years of discussions and work on Bell's Theorem, quantum interferometry and quantum information. Most of the present work was done during author's multiple visits in Vienna, and Singapore (many thanks to profs Ekert, Kwek and Oh). Special thanks are for David Mermin with whom the author intensively discussed the earlier 
versions of this paper, and especially for finding a mistake in one of the central proofs in the first version.

The author is supported by a Professorial Subsidy of FNP. The work is part of MNiI
grant No PBZ-MIN-008/ P03/ 2003.

\end{document}